\documentclass[12pt,a4paper]{article}
\usepackage{amsfonts,latexsym}
\usepackage{amsmath,amssymb}
\usepackage{graphicx,color}

\renewcommand{\thefootnote}{\fnsymbol{footnote}}

\begin{document}

\vspace{12mm}

\begin{center}
{{{\Large {\bf Scale-invariant spectrum of Lee-Wick model in de Sitter
spacetime}}}}\\[10mm]

{Yun Soo Myung\footnote{e-mail address: ysmyung@inje.ac.kr} and Taeyoon Moon\footnote{e-mail address: tymoon@inje.ac.kr}}\\[8mm]

{Institute of Basic Sciences and Department  of Computer Simulation, Inje University Gimhae 621-749, Korea\\[0pt]}

\end{center}
\vspace{2mm}

\begin{abstract}
We obtain a scale-invariant spectrum from the Lee-Wick model in de
Sitter spacetime. This model is a fourth-order scalar theory whose
mass parameter is determined by $M^2=2H^2$. The Harrison-Zel'dovich
 scale-invariant
 spectrum is obtained by Fourier transforming the propagator in
position space as well as by computing the power spectrum directly.
It shows clearly that the LW scalar theory provides a truly
scale-invariant spectrum in whole de Sitter, while  the massless
scalar propagation in de Sitter shows a scale-invariant spectrum  in
the superhorizon region only.

\end{abstract}
\vspace{5mm}



\vspace{1.5cm}

\hspace{11.5cm}{Typeset Using \LaTeX}
\newpage
\renewcommand{\thefootnote}{\arabic{footnote}}
\setcounter{footnote}{0}

\section{Introduction}

 The Lee-Wick (LW) model \cite{Lee:1969fy,Lee:1970iw} of a fourth-order
 derivative
scalar theory with  $\phi$ has provided a cosmological bounce which
could avoid the  singularity and give a scale-invariant
spectrum~\cite{Cai:2008qw}. Introducing an auxiliary field (LW
scalar)  and  a normal scalar = $\phi+$ LW
scalar~\cite{Grinstein:2007mp}, the fourth-order Lagrangian can be
expressed in terms of two second-order Lagrangians. Here the kinetic
and mass terms of the LW scalar have the opposite sign when compared
with those  for the normal scalar. Since the LW scalar is  a ghost
scalar, it provides a bouncing solution.  Perturbations  of normal
scalar generated  in the contracting phase have survived during
bouncing and have led to a scale-invariant spectrum in the expanding
phase~\cite{Wands:1998yp,Finelli:2001sr}. Thus, the LW model is
considered as  a possible alternative to inflationary cosmology.

It is well-known that the power spectrum of a massless minimally
coupled scalar (mmc) in de Sitter (dS) spacetime takes  the form of
$(H/2\pi)^2[1+k^2/(a^2H^2)]$ which reduces to the
Harrison-Zel'dovich (HZ) scale-invariant spectrum of $(H/2\pi)^2$ in
the superhorizon region of $k \ll aH$~\cite{Baumann:2009ds}. On the
other hand, the quantization of a mmc scalar field in dS spacetime
has a nontrivial problem due to the appearance of IR divergence in
compared to a massive minimally coupled scalar~\cite{Allen:1987tz}.
They then propose to trade dS SO(1,4) invariance for a smaller SO(4)
invariance.  Actually, the construction of  the mmc scalar Green
function in coordinate-space has remained a matter of controversy
and has been considered as a subject of debate for past decades.
After subtracting the divergent term, however, one has gotten a
renormalized Green function (propagator) $G_{\rm mmc}(Z(x,x'))$ with
a tractable drawback \cite{Bros:2010wa,Antoniadis:2011ib} where
$Z(x,x')$ is dS SO(1,4) invariant distance.  Using Cesaro-summation
method to tame a divergent Fourier transform properly, Youssef has
recently recovered the original form of power spectrum but not a
scale-invariant (equal amplitude on all scales) spectrum in whole dS
region~\cite{Youssef:2012cx}.

At this stage, we ask an important question ``what kind of a scalar
model could provide a truly scale-invariant power spectrum in whole
dS evolution". The answer is that it would  be the LW model with
mass parameter $M^2=2H^2$. In this case, we will not introduce the
auxiliary field method (normal and LW scalars to lower the LW model
to a second-order theory with two scalars). Instead, one uses the
Ostrogradski formalism~\cite{Mannheim:2004qz,de Urries:1998bi} and
their equivalence was proved in appendix B of
Ref.\cite{Chen:2013aha}.  The LW operator
$\Delta_4=-\bar{\nabla}^2(-\bar{\nabla}^2+2H^2)$ is a conformally
covariant fourth-order  operator  since it transforms
$\Delta_4=e^{-4\sigma}\tilde{\Delta}_4$ under the conformal
transformation of $g_{\mu\nu}\to e^{2\sigma}\tilde{g}_{\mu\nu}$  in
dS spacetime~\cite{Mazur:2001aa}. Furthermore, its propagator takes
the form $\tilde{D}(Z(x,x'))=[G_{\rm mmc}(Z(x,x'))-G_{\rm
mcc}(Z(x,x'))]=-\frac{H^2}{8\pi^2}\ln[1-Z(x,x')]$ where $G_{\rm
mcc}(Z(x,x'))$ represents the propagator of a massless conformally
coupled (mcc) scalar.  It turns out that taking Cesaro-summation
method to define a divergent Fourier transform of
$\tilde{D}(Z(x,x'))$
 leads to an exactly  scale-invariant spectrum of $(H/2\pi)^2$ without scale-dependence $k$.
 This implies that the HZ scale-invariant
 spectrum corresponds to a logarithmic zero conformal weight distribution in
 the coordinate-space S$^2$ angular directions on the sky~\cite{Antoniadis:2011ib}. In this case, $Z(x,x')=\hat{n}\cdot\hat{n}'$
 is the  cosine of the angle between the two direction vectors on the sky viewed from the origin where the radiation appears to originate.

 In this work, we compute the power spectrum of the LW model directly  by
 employing the quantization scheme of Pais-Uhlenbeck fourth-order oscillator \cite{Mannheim:2004qz} and taking the
 Bunch-Davies vacuum. As was expected, we obtain the same scale-invariant
 spectrum of $(H/2\pi)^2$. This implies clearly that a scale-invariant
 spectrum preserves  dS SO(1,4) invariance.

\section{Einstein-Lee-Wick gravity}

Let us first consider the Einstein-Lee-Wick (ELW) gravity whose
action is given by
\begin{eqnarray} \label{elw}
S_{\rm ELW}=S_{\rm E}+S_{\rm LW}=\int d^4x
\sqrt{-g}\Big[\Big(\frac{R}{2\kappa}-\Lambda\Big)+\frac{1}{2}\Big(\phi\nabla^2\phi-\frac{1}{M^2}(\nabla^2\phi)^2\Big)\Big],
\end{eqnarray}
where $\kappa=8\pi G=1/M^2_{\rm P}$, $M_{\rm P}$ being the reduced
Planck mass and $M^2$ is a mass parameter determined to be  $2H^2$.
Here, $S_{\rm E}$ denotes the action for the Einstein gravity with
positive cosmological constant, whereas $S_{\rm LW}$ is  the LW
scalar action which differs slightly from the the LW standard action
including a mass term $m^2\phi^2$ and potential
$\frac{g}{3!}\phi^3$~\cite{Grinstein:2007mp}.

Varying the  action (\ref{elw}) with respect to the metric tensor
$g_{\mu\nu}$ leads to  the Einstein equation
\begin{equation} \label{einseq}
G_{\mu\nu} +\kappa \Lambda g_{\mu\nu}=\kappa T_{\mu\nu},
\end{equation}
where  the energy-momentum tensor takes the form
\begin{eqnarray}
T_{\mu\nu}&=&\nabla_{\mu}\phi\nabla_{\nu}\phi+\frac{1}{2}g_{\mu\nu}\nabla_{\rho}\phi\nabla^{\rho}\phi\nonumber\\
&&-\frac{1}{M^2}\Big[2\nabla_{\mu}(\nabla^2\phi)\nabla_{\nu}\phi +
g_{\mu\nu}\nabla_{\rho}(\nabla^2\phi)\nabla^{\rho}\phi-\frac{1}{2}g_{\mu\nu}\nabla^2\phi\nabla^2\phi\Big].
\end{eqnarray}
On the other hand, the scalar  equation for the action (\ref{elw})
is given by
\begin{equation} \label{scalar-eq}
\nabla^2\phi-\frac{1}{M^2}\nabla^2\nabla^2\phi=0\to
-\frac{1}{M^2}\nabla^2(\nabla^2-M^2)\phi=0.
\end{equation}
 A solution of dS spacetime  to Eq.(\ref{einseq}) can be easily found  when one chooses  the vanishing scalar
\begin{equation}
\bar{R}=4\kappa \Lambda,~~\bar{\phi}=0.
\end{equation}
In this case, the Riemann and Ricci tensors can be written by
\begin{equation}
\bar{R}_{\mu\nu\rho\sigma}=H^2(\bar{g}_{\mu\rho}\bar{g}_{\nu\sigma}-\bar{g}_{\mu\sigma}\bar{g}_{\nu\rho}),~~\bar{R}_{\mu\nu}=3H^2\bar{g}_{\mu\nu}
\end{equation}
with  Hubble constant $H^2=\kappa \Lambda/3$. Also, the dS spacetime
can be represented by  introducing  either cosmic time $t$ or
conformal time $\eta$ as
\begin{eqnarray} \label{deds1}
ds^2_{\rm dS}=\bar{g}_{\mu\nu}dx^\mu
dx^\nu&=&-dt^2+a^2(t)\delta_{ij}dx^idx^j\\
\label{deds2}&=&a(\eta)^2[-d\eta^2+\delta_{ij}dx^idx^j],
\end{eqnarray}
where $a(t)$ and $a(\eta)$ are cosmic and  conformal  scale factors
expressed by
\begin{eqnarray}
a(t)=e^{Ht},~~a(\eta)=-\frac{1}{H\eta}.
\end{eqnarray}
 During the de Sitter stage, $a$ goes from small to a very
large value like $a_f / a_i\simeq 10^{30}$ which implies that the
conformal time $\eta=-1/aH(z=-k\eta)$ runs from
$-\infty(\infty)$[the infinite past] to $0^-(0)$ [the infinite
future].
 The dS SO(1,4) invariant distance between two spacetime points $x^\mu$ and $x'^{\mu}$ is given by
\begin{eqnarray}
&&Z(x,x')=\frac{1}{2}\Big[1-\frac{H^2e^{H(t+t')}}{2}|{\bf{x}}-{\bf{x}'}|^2+\cosh[H(t-t')]\Big],
\label{id1}
\\
\label{id2}
&&Z(x,x')=1-\frac{1}{4\eta\eta'}\Big[-(\eta-\eta')^2+|{\bf{x}}-{\bf{x}'}|^2\Big],
\end{eqnarray}
where the former  is the distance when using (\ref{deds1}), while
the latter is the distance for (\ref{deds2}).

\section{Scalar propagation in dS spacetime}

To investigate the cosmological perturbation around the dS spacetime
(\ref{deds2}), we might  choose the Newtonian gauge as $B=E=0 $ and
$\bar{E}_i=0$. Under this gauge, the corresponding perturbed metric
with transverse-traceless tensor $\partial_ih^{ij}=h=0$ and
perturbed scalar can be written as
\begin{eqnarray} \label{so3ds}
ds^2&=&a(\eta)^2\Big[-(1+2\Psi)d\eta^2+2\Psi_i d\eta
dx^{i}+\Big\{(1+2\Phi)\delta_{ij}+h_{ij}\Big\}dx^idx^j\Big],\\
\phi&=&
\bar{\phi}+\varphi\label{phip}.
\end{eqnarray}
Now we linearize the Einstein equation (\ref{einseq}) around the dS
background to obtain  the cosmological perturbed equations. It is
known that the tensor perturbation  is decoupled from scalars and
its equation becomes
\begin{eqnarray}
\delta R_{\mu\nu}(h)-3H^2h_{\mu\nu}=0 \to
\bar{\nabla}^2h_{ij}=0.\label{heq}
\end{eqnarray}
We mention briefly how do two scalars $\Psi$ and $\Phi$, and a
vector $\Psi_i$ go on. The linearized Einstein equation requires
$\Psi=-\Phi$ which was  used to define the comoving curvature
perturbation in the slow-roll inflation and thus, they are not
physically propagating modes. During the dS inflation, no coupling
between $\{\Psi,\Phi\}$ and $\varphi$ occurs because of
$\bar{\phi}=0$. Lastly, the vector is also a non-propagating mode in
the ELW theory because it has no  kinetic term.

On the other hand, the linearized scalar equation is given by
\begin{equation} \label{s-eq}
\bar{\nabla}^2(\bar{\nabla}^2-M^2)\varphi=0.
\end{equation}
Hereafter we choose $M^2$ to be $2H^2$ to get a mcc  scalar sector.

In order to find the solution to the linearized fourth-order
equation (\ref{s-eq}) in the  whole range of $\eta$, we decompose
(\ref{s-eq}) into two second-order equations
\begin{eqnarray}
\bar{\nabla}^2\varphi^{({\rm mmc})}&=&0,\label{eeq}\\
(\bar{\nabla}^2-2H^2)\varphi^{({\rm mcc})}&=&0\label{meq},
\end{eqnarray}
where $\varphi=\varphi^{({\rm mmc})}+\varphi^{({\rm
mcc})}\equiv\varphi^{(1)}+\varphi^{(2)}$. This is always possible to
occur for a direct product form of fourth-order equation as in
(\ref{s-eq}). Expanding $\varphi^{(i)}$ in terms of Fourier modes
$\phi^{(i)}_{\bf k}(\eta)$
\begin{eqnarray}\label{sfour}
\varphi^{(i)}(\eta,{\bf x})=\frac{1}{(2\pi)^{\frac{3}{2}}}\int
d^3{k}~\phi^{(i)}_{\bf k}(\eta)e^{i{\bf k}\cdot{\bf x}},
\end{eqnarray}
equations (\ref{eeq}) and (\ref{meq}) become
\begin{eqnarray}\label{s-eq2}
\Bigg[\frac{d^2}{d z^2}-\frac{2}{z}\frac{d}{d
z}+1\Bigg]\phi^{({1})}_{\bf
k}&=&0,\label{pmsol}\\
\Bigg[\frac{d^2}{dz^2}-\frac{2}{z}\frac{d}{d
z}+1+\frac{2}{z^2}\Bigg]\phi^{(2)}_{\bf k}&=&0\label{pmmsol}
\end{eqnarray}
with  $z=-\eta k$. Solutions to (\ref{pmsol}) and (\ref{pmmsol}) are
easily  found to be
\begin{eqnarray}
\phi_{\bf k}^{(1)}&=&{\cal C}_1(i+z)e^{iz},\label{pmsol1}\\
\phi_{\bf k}^{(2)}&=&{\cal C}_2ize^{iz},\label{pmmsol1}
\end{eqnarray}
where ${\cal C}_{1}$ and ${\cal C}_{2}$ are constants to be
determined.

\section{Propagator in de Sitter}

We wish to find the power spectrum of perturbed scalar  by making
Fourier transform of propagator in dS spacetime.   First of all, we
introduce the LW operator defined
by~\cite{Mazur:2001aa,Mottola:2010gp}
\begin{eqnarray}
\Delta_4&=&\bar{\nabla}^4+2R^{\mu\nu}\bar{\nabla}_\mu\bar{\nabla}_\nu-\frac{2}{3}R\bar{\nabla}^2+\frac{1}{3}(\bar{\nabla}^\mu
R)\bar{\nabla}_\mu \nonumber \\
&\xrightarrow[\rm dS]{}&-\bar{\nabla}^2(-\bar{\nabla}^2+2H^2)
\end{eqnarray}
which is a conformally covariant fourth-order operator because it
transforms $\Delta_4=e^{-4\sigma}\tilde{\Delta}_4$ under a rescaling
of metric $g_{\mu\nu}\to e^{2\sigma}\tilde{g}_{\mu\nu}$ in dS
spacetime. Accordingly, $\sqrt{-g}\Delta_4$ is a conformally
invariant operator.  The propagator is given by the inverse of
$\Delta_4$ as~\cite{Antoniadis:2011ib}
\begin{equation} \label{4thp}
D(Z(x,x'))=\frac{1}{2H^2}\Big[\frac{1}{-\bar{\nabla}^2}-\frac{1}{-\bar{\nabla}^2+2H^2}\Big]
=\frac{1}{2H^2}[G_1(Z(x,x'))-G_2(Z(x,x'))]
\end{equation}
where the propagators of  1(mmc) and 2(mcc) scalar in dS spacetime
are given by
\begin{equation}\label{prop}
G_1(Z(x,x'))=\frac{H^2}{(4\pi)^2}\Big[\frac{1}{1-Z}-2\ln(1-Z)+c_0\Big],~~G_2(Z(x,x'))=\frac{H^2}{(4\pi)^2}\frac{1}{1-Z}.
\end{equation}
On the other hand, upon choosing the Bunch-Davies vacuum,  the
propagator of a massive minimally coupled scalar is given by the
hypergeometric function~\cite{Chernikov:1968zm}
\begin{equation} \label{green}
G_0(Z,m^2)=\frac{H^2}{(4\pi)^2}\Gamma(\triangle_+)\Gamma(\triangle_-)~_2F_1(\triangle_+,\triangle_-,2;Z(x,x'))
\end{equation}
with $\triangle_\pm=\frac{3}{2}\pm
\sqrt{\frac{9}{4}-\frac{m^2}{H^2}}$ for $0<m^2\le \frac{9}{4}H^2$.
For a mmc ($m^2=0,\triangle_+=3,\triangle_-=0$) scalar, the
quantization of a mmc scalar field in dS spacetime has a nontrivial
problem due to the appearance of IR divergence ($\Gamma(0)$) in
compared to a massive minimally coupled scalar~\cite{Allen:1987tz}.
After subtracting the divergent term, one got a renormalized
propagator $G_1(Z(x,x'))$ in (\ref{prop}) with a tractable
drawback~\cite{Bros:2010wa}.  In the case of a mcc scalar with
$m^2=2H^2(\triangle_+=2,\triangle_-=1)$, the corresponding
propagator is given by $G_2(Z(x,x'))$~\cite{Higuchi:2009ew}.

Plugging (\ref{prop}) into (\ref{4thp}), its propagator takes the
form \begin{equation}
D(Z(x,x'))=\frac{1}{16\pi^2}\Big(-\ln[1-Z(x,x')]+\frac{c_0}{2}\Big)
\end{equation}
 which is a pure logarithm up to an arbitrary additive constant
 $c_0$.
Since our propagator relation is read off  from (\ref{scalar-eq})
\begin{equation}
\label{o4thp}
\tilde{D}(Z(x,x'))=\Big[\frac{1}{-\bar{\nabla}^2}-\frac{1}{-\bar{\nabla}^2+2H^2}\Big]
=[G_1(Z(x,x'))-G_2(Z(x,x'))],
\end{equation}
it takes the form \label{o4thp}
\begin{equation}
\tilde{D}(Z(x,x'))=-\frac{H^2}{8\pi^2}\ln[1-Z(x,x')]
\end{equation}
with $c_0=0$ for simplicity. The power spectrum is then formally
given by \begin{equation} {\cal P}=\frac{1}{(2\pi)^3}\int d^3{r}
~4\pi k^3 \tilde{D}(Z({\bf x},t;{\bf x}',t))e^{-i{\bf k}\cdot{\bf
r}},~~{\bf r}={\bf x}-{\bf x}'.
\end{equation}

 It turns out that making use of
Cesaro-summation method to compute a divergent Fourier transform of
$\tilde{D}(Z(x,x'))$~\cite{Youssef:2012cx}, we
 obtain  an exactly  scale-invariant spectrum
 \begin{equation}
{\cal P}=\Big(\frac{H}{2\pi}\Big)^2
\end{equation}
without scale-dependence $k$.

\section{Power spectra}

In order to find power spectrum for a scalar perturbation in the ELW
gravity, we  rewrite the fourth-order bilinear action $\delta S_{\rm
LW}$ (\ref{elw}) as the second-order bilinear action
 by using the Ostrogradski's formalism for scalar~\cite{Mannheim:2004qz,de Urries:1998bi} and tensor~\cite{Deruelle:2012xv,Myung:2014jha} as
\begin{eqnarray}
&&\hspace*{-2.3em}\delta S^2_{\rm LW}=\frac{1}{2}\int
d^4x\Big[-a^2(\alpha^2+\partial_i\varphi\partial^i\varphi)
-\frac{1}{2H^2}\Big((\alpha')^2-2\partial_i\alpha\partial^i\alpha
+\partial^2\varphi\partial^2\varphi\nonumber\\
&&\hspace*{8em}+4aH\alpha\alpha'-4aH\alpha\partial^2\varphi\Big)
+2\beta(\alpha-\varphi')\Big]\label{slw1},
\end{eqnarray}
where $\alpha\equiv \varphi'$ is a new field,
$\partial^2=\partial_i\partial^i$, and $\beta$ is a Lagrange
multiplier. Here the prime ($'$) denotes differentiation with
respect to $\eta$. From (\ref{slw1}), we define the conjugate
momenta as
\begin{eqnarray}\label{conjm}
\pi_{\varphi}=\frac{1}{2H^2}\Big(\varphi'''-2\partial^2\varphi'+2aH\partial^2\varphi\Big),
~~~\pi_{\alpha}=-\frac{1}{2H^2}(\varphi''+2aH\varphi').
\end{eqnarray}
Then, the canonical quantization is accomplished by imposing
commutation relations as follows:
\begin{eqnarray}\label{comr}
[\hat{\varphi}(\eta,{\bf x}),~\hat{\pi}_{\varphi}(\eta,{\bf
x}^{\prime})]=i\delta({\bf x}-{\bf
x}^{\prime}),~~~[\hat{\alpha}(\eta,{\bf
x}),~\hat{\pi}_{\alpha}(\eta,{\bf x}^{\prime})]=i\delta({\bf
x}-{\bf x}^{\prime}).
\end{eqnarray}
The field  operator $\hat{\varphi}$ can be expanded in Fourier modes
as
\begin{eqnarray}\label{phiex}
\hat{\varphi}(\eta,{\bf x})=\frac{1}{(2\pi)^{\frac{3}{2}}}\int
d^3k\Big[\Big(\hat{a}_{\bf k}\phi_{\bf k}^{(1)}(\eta)+\hat{b}_{\bf
k}\phi_{\bf k}^{(2)}(\eta)\Big)e^{i{\bf k}\cdot{\bf x}}+~{\rm
h.c.}\Big].
\end{eqnarray}
We  also obtain  the momentum operator $\hat{\pi}_{\varphi}$ by
substituting (\ref{phiex}) into (\ref{conjm})
\begin{eqnarray}\label{phiex1}
&&\hspace*{-3em}\hat{\pi}_{\varphi}(\eta,{\bf
x})=\frac{1}{(2\pi)^{\frac{3}{2}}}\frac{1}{2H^2}\int
d^3k\Big[\Big(\hat{a}_{\bf k}\Big\{\Big(\phi_{\bf
k}^{(1)}(\eta)\Big)'''+2k^2\Big(\phi_{\bf
k}^{(1)}(\eta)\Big)'-2aHk^2\phi_{\bf k}^{(1)}(\eta)\Big\}e^{i{\bf
k}\cdot{\bf
x}}\nonumber\\
&&\hspace*{3.8em}+~\hat{b}_{\bf k}\Big\{\Big(\phi_{\bf
k}^{(2)}(\eta)\Big)'''+2k^2\Big(\phi_{\bf
k}^{(2)}(\eta)\Big)'-2aHk^2\phi_{\bf k}^{(2)}(\eta)\Big\}e^{i{\bf
k}\cdot{\bf x}}\Big)+{\rm h.c.}\Big].
\end{eqnarray}
Similarly,  $\hat{\alpha}(\equiv\hat{\varphi}')$ operator  and its
momentum operator $\hat{\pi}_{\alpha}$ (\ref{conjm}) can be
expressed as
\begin{eqnarray}\label{alex}
&&\hat{\alpha}(\eta,{\bf x})=\frac{1}{(2\pi)^{\frac{3}{2}}}\int
d^3k\Big[\Big\{\hat{a}_{\bf k}\Big(\phi_{\bf
k}^{(1)}(\eta)\Big)'+\hat{b}_{\bf k}\Big(\phi_{\bf
k}^{(2)}(\eta)\Big)'\Big\}e^{i{\bf k}\cdot{\bf x}}+~{\rm h.c.}\Big],\\
&&\hat{\pi}_{\alpha}(\eta,{\bf
x})=-\frac{1}{(2\pi)^{\frac{3}{2}}}\frac{1}{2H^2}\int
d^3k\Big[\hat{a}_{\bf k}\Big\{\Big(\phi_{\bf
k}^{(1)}(\eta)\Big)''+2aH\Big(\phi_{\bf
k}^{(1)}(\eta)\Big)'\Big\}e^{i{\bf k}\cdot{\bf
x}}\nonumber\\
&&\hspace*{7.8em}+~\hat{b}_{\bf k}\Big\{\Big(\phi_{\bf
k}^{(2)}(\eta)\Big)''+2aH\Big(\phi_{\bf
k}^{(2)}(\eta)\Big)'\Big\}e^{i{\bf k}\cdot{\bf x}}+{\rm
h.c.}\Big].\label{alex1}
\end{eqnarray}
Substituting (\ref{phiex})-(\ref{alex1}) into (\ref{comr}) leads to
the commutation relations and Wronskian conditions:
\begin{eqnarray}
&&\hspace*{-2em}[\hat{a}_{\bf k},~\hat{a}_{\bf k^{\prime}}^{\dag}]=\delta({\bf k}-{\bf
k}^{\prime}),~~~~~~[\hat{b}_{\bf k},~\hat{b}_{\bf
k^{\prime}}^{\dag}]=-\delta({\bf
k}-{\bf
k}^{\prime}),\label{comrell}\\
\nonumber\\
&&\hspace*{-2em}\Big[\phi_{\bf k}^{(1)}\Big\{\Big(\phi_{\bf
k}^{*(1)}(\eta)\Big)'''+2k^2\Big(\phi_{\bf
k}^{*(1)}(\eta)\Big)'-2aHk^2\phi_{\bf
k}^{*(1)}(\eta)\Big\}\nonumber\\
&&-\phi_{\bf k}^{(2)}\Big\{\Big(\phi_{\bf
k}^{*(2)}(\eta)\Big)'''+2k^2\Big(\phi_{\bf
k}^{*(2)}(\eta)\Big)'-2aHk^2\phi_{\bf
k}^{*(2)}(\eta)\Big\}\Big]-c.c.=i2H^2,\label{wcona}\\
\nonumber\\
&&\hspace*{2em}\Big[\Big(\phi_{\bf
k}^{(1)}\Big)'\Big\{\Big(\phi_{\bf
k}^{*(1)}(\eta)\Big)''+2aH\Big(\phi_{\bf
k}^{*(1)}(\eta)\Big)'\Big\}\nonumber\\
&&\hspace*{5em}-\Big(\phi_{\bf k}^{(2)}\Big)'\Big\{\Big(\phi_{\bf
k}^{*(2)}(\eta)\Big)''+2aH\Big(\phi_{\bf
k}^{*(2)}(\eta)\Big)'\Big\}\Big]-c.c.=-i2H^2.\label{wconb}
\end{eqnarray}
 We note that inspired by quantization of
the Pais-Uhlenbeck fourth-order oscillator~\cite{Mannheim:2004qz},
two mode operators ($\hat{a}_{\bf k},\hat{b}_{\bf k}$) are necessary
to take into account of fourth-order scalar theory. The unusual
commutator for ($\hat{b}_{\bf k},\hat{b}_{\bf k^{\prime}}^{\dag}$)
reflects that the LW model contains the ghost state
scalar~\cite{Chen:2013aha}. Before we proceed, we remind the reader
that $\phi_{\bf k}^{(1)}$ and $\phi_{\bf k}^{(2)}$ are given by
(\ref{pmsol1}) and (\ref{pmmsol1}), respectively. Making use of the
Wronskian conditions (\ref{wcona}) and (\ref{wconb}) determine these
solutions completely  as
\begin{eqnarray}
\phi_{\bf k}^{(1)}&=&\frac{H}{\sqrt{2k^3}}(i+z)e^{iz},\label{p1sol}\\
\phi_{\bf k}^{(2)}&=&\frac{H}{\sqrt{2k^3}}ize^{iz},\label{p2sol}
\end{eqnarray}
which implies that  $|{\cal C}_1|^2=|{\cal C}_2|^2=H^2/(2k^3)$.  One
 checks easily that  solutions (\ref{p1sol}) and (\ref{p2sol}) also
satisfy the initial condition when choosing  the Bunch-Davies vacuum
$|0\rangle$ in the subhorizon limit ($z\to \infty$) of Eqs.
(\ref{pmsol}) and (\ref{pmmsol}).

On the other hand, the power spectrum of the scalar  is defined
by~\cite{Baumann:2009ds}
\begin{eqnarray}\label{pow}
\langle0|\hat{\varphi}(\eta,{\bf x})\hat{\varphi}(\eta,{\bf
x^{\prime}})|0\rangle=\int d^3k\frac{{\cal P}_{\varphi}}{4\pi
k^3}e^{i{\bf k}\cdot({\bf x}-{\bf x^{\prime}})},
\end{eqnarray}
which leads to the HZ scale-invariant spectrum
\begin{eqnarray}
{\cal P}_{\rm \varphi}&=&\frac{k^3}{2\pi^2}\left(\Big|\phi_{\bf
k}^{(1)}\Big|^2-\Big|\phi_{\bf k}^{(2)}\Big|^2\right)\label{powmM}\\
&=&\left(\frac{H}{2\pi}\right)^2\Big[1+\frac{k^2}{(aH)^2}-\frac{k^2}{(aH)^2}\Big]=\left(\frac{H}{2\pi}\right)^2.
\end{eqnarray}
 Here we used the Bunch-Davies vacuum state by imposing $\hat{a}_{\bf
k}|0\rangle=0 $ and $\hat{b}_{\bf k}|0\rangle=0$, and the minus sign
($-$) in (\ref{powmM}) appears when using the  unusual commutation
relation for ($\hat{b}_{\bf k},\hat{b}_{\bf k^{\prime}}^{\dag}$).

Finally, by comparing (\ref{heq}) with (\ref{eeq}), the tensor power
spectrum is given by
\begin{equation} \label{h-spec}
{\cal P}_{\rm h}=2\times \Big(\frac{2}{M_{\rm P}}\Big)^2\times {\cal
P}_{\varphi^{(1)}}=\frac{2H^2}{\pi^2 M^2_{\rm
P}}\Big[1+\frac{k^2}{(aH)^2}\Big],
\end{equation}
where ${\cal P}_{\varphi^{(1)}}$ is the spectrum for the mmc scalar.

\section{Discussions}

First of all, the LW model is regarded as a simple fourth-order
scalar theory. In this work, we have derived the Harrison-Zel'dovich
scale-invariant spectrum by Fourier transforming the
coordinate-space renormalized propagator of the LW model with mass
parameter $M^2=2H^2$ in dS spacetime.  In deriving it, we have used
Cesaro-summation technique.

 Also, the same scale-invariant power spectrum have been
found directly by employing the quantization scheme for a
Pais-Uhlenbeck  fourth-order oscillator and taking the Bunch-Davies
vacuum for dS spacetime. This shows that the scale-invariant
spectrum comes out, while preserving dS SO(1,4) symmetry. In this
computation, we have used the Ostrogradski's formalism instead of
the auxiliary formalism because we want to derive the power spectrum
of  a single scalar satisfying a fourth-order equation, but not for
the normal and LW scalars satisfying second-order equations,
respectively.

 As was shown in (\ref{h-spec}), the tensor spectrum is not
 scale-invariant in whole dS space but it is scale-invariant in the
 superhorizon region of $k \ll aH$. Hence, we have to find the
 corresponding tensor theory  that is similar to the Lee-Wick scalar theory.
We propose that it might be the massive conformal gravity
 with an appropriate choice of  mass
parameter~\cite{Faria:2013hxa,Myung:2014aia}, which will be explored
elsewhere.

Consequently, the HZ scale-invariant spectrum of scalar is not given
by a massless scalar theory but  by LW scalar theory in dS
spacetime. This means that the original dS SO(1,4) symmetry
preserves in computing propagator (power spectrum) of LW scalar
theory.  Also, it is worth noting that the massless scalar operator
($\sqrt{-g}\nabla^2$) is not conformally invariant, while the LW
operator is conformally invariant ($\sqrt{-g}\Delta_4^2\to
\sqrt{-\tilde{g}}\tilde{\Delta}_4$) under the conformal
transformation of $g_{\mu\nu} \to e^{2\sigma}\tilde{g}_{\mu\nu}$.

 \vspace{0.25cm}

{\bf Acknowledgement}

\vspace{0.25cm}
 This work was supported by the National
Research Foundation of Korea (NRF) grant funded by the Korea
government (MEST) (No.2012-R1A1A2A10040499).

\end{document}